# The Critical Effect of Hydration on the Resonant Signatures of THz Biospectroscopy

**E.R. Brown and W-D. Zhang**

Department of Physics, Wright State University, Dayton, OH 45435    USA

elliott.brown@wright.edu, (001) 937-775-4902

**Abstract** Here we present an original study of the effect of hydration on THz absorption signatures in biomolecules (lactose monohydrate and biotin) and bioparticles (Bacillus thuringiensis and Bacillus cereus spores). We observe a "redshift" in center frequency with increasing hydration in all samples, consistent with Lorentzian-oscillator behavior. But the effect of hydration on linewidth is ambiguous, sometimes increasing the linewidth (consistent with Lorentzian behavior) and sometimes decreasing the linewidth.

## 1.1  Introduction

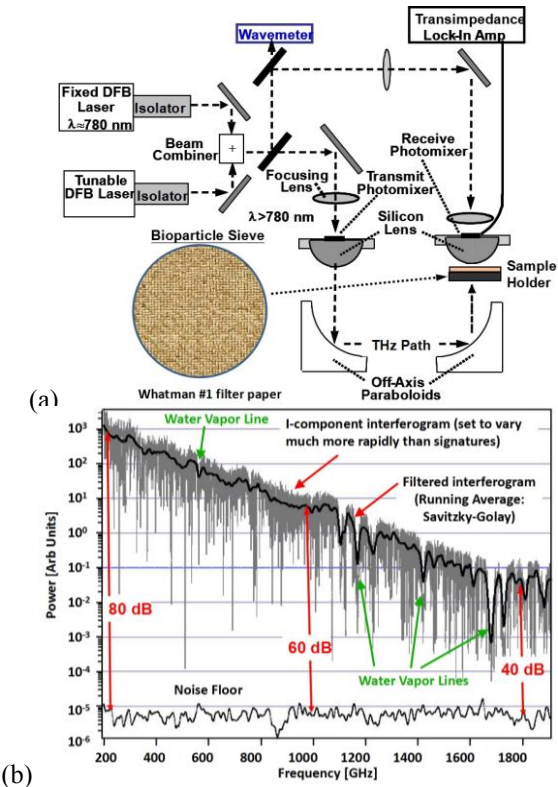

Fig. 1. (a) Photomixing spectrometer block diagram and location of biosamples, such as the Bacillus spores embedded in filter paper. (b) Typical performance of spectrometer from ~0.2 to 1.9 THz.

Terahertz (THz) radiation lies in the range 300-10000 GHz, which currently is beyond the upper-frequency limit of wireless-communications and radar systems worldwide. It has been investigated for a variety of applications such as radio astronomy [1], security imaging [2], materials characterization [3], and medical diagnostics [4]. And of course, these applications benefit from the continual evolution of solid-state electronics and electromagnetics, such as resonant-tunneling oscillators up to 1.2 THz [5], amplifiers up to 1.0 THz [6], and frequency-multipliers and mixers up to roughly 2.0 THz [7]. Phenomenologically, MMW and THz radiation oscillate with a period comparable to the ps-timescale of collective vibrations in biomolecules and bioparticles, and also to the characteristic orientation times of dipolar liquid-water molecules. Because of this both are absorbed strongly by soft tissue of all sorts, but the collective interaction of molecules and particles can be more resonant with respect to excitation frequency. In principle, this makes it possible to distinguish the biomolecular excitation from water absorption, which generally causes heating. Heating in biomaterials is complicated and potentially confusing since it creates its own strong bioeffeccts.

For many years THz spectroscopy has been touted as a complement to FTIR and microwave vapor-phase spectroscopies, providing information about vibrational resonances from the tertiary structure of biomolecules and bioparticles rather than their interatomic vibrations or rotations. The challenge in THz spectroscopy has been achieving adequate selectivity and specificity needed to distinguish these signatures from various forms of "clutter", such as standing waves in sample holders and scattering effects (e.g., speckle) in grainy materials. Another challenge is measuring biosignatures "in-situ", meaning in hydrated media for biosamples. The strong absorption from liquid water makes spectroscopy virtually impossible for most living organisms *in vivo*, and even most aqueous solutions. The lack of powerful THz sources (i.e., average output ≥ 1 mW) exacerbates the problem since traditional methods for clutter mitigation, such as the placement of attenuators in the signal path to suppress standing waves, tend to reduce the THz signal-to-noise ratio deleteriously.

In this work we show that the hydration level for THz signatures is "critical", meaning that it must lie in a certain range for signatures to be readily measured. For polar molecules having a strong enough vibrational resonance, such as lactose monohydrate, the lower end of the range extends to zero and the upper end to roughly 20% hydration by mass. For non-polar molecules or bioparticles, such as Bacillus spores, the lower end of the range rises above zero to roughly 10% hydration, but the upper end remains at about 20%. In addition, hydration usually causes a reduction (i.e., "red shift") in the signature center frequency which will be described later in the document.

## 1.2 Experimental Methods and Samples

The spectroscopic experiments were carried out with a high-resolution frequency-domain, coherent (photomixing) spectrometer like those described in References [8] and [9]. The spectra were measured in stepped-frequency mode at 1.0 GHz steps and 0.1 s integration time. The typical background spectra and noise floor are shown in Fig. 1(a) where the tuning bandwidth is seen to be at least 1.7 THz, and the dynamic range is ≈80 dB at the low end of the scan (200 GHz) and ≈40 dB at the high end (1.9 THz). The frequency range is determined mostly by the limited temperature tuning of the DFB diode lasers used in the spectrometer. For the present experiments, the frequency resolution is determined by the 1.0 GHz steps, although the instantaneous THz tone is much narrower than this.

For the present study we chose two biomolecular samples and two bioparticle samples, all known from previous experience to have strong signatures in the THz region and within the operating range of our photomixing spectrometer. The biomolecular samples were lactose monohydrate and biotin, each having the chemical compositions shown in Fig. 2(a) and (b), respectively. The former is a crystallized form of milk sugar (monoclinic structure), and the latter is crystallized vitamin B7 (rhombohedral structure). Both molecular crystals have built-in polarity, are hydrophylic, and appear macroscopically as white powders with submicron grains. The lactose and biotin samples were mounted in metallic rings and encapsulated with thin (~12 um) polycellulose windows to slow the rate of evaporation. Two identical samples were made for each biomolecular type and loaded to the same approximate level of hydration, starting at between 25% and 30% *gravimetric* (accurate to ≈0.1%). One of the two was left on a precision balance, and the other mounted in the THz spectrometer. Then THz spectra were taken roughly every 15 minutes, corresponding to a typical drop of hydration on the balance-sample of ≈2% per spectrum.

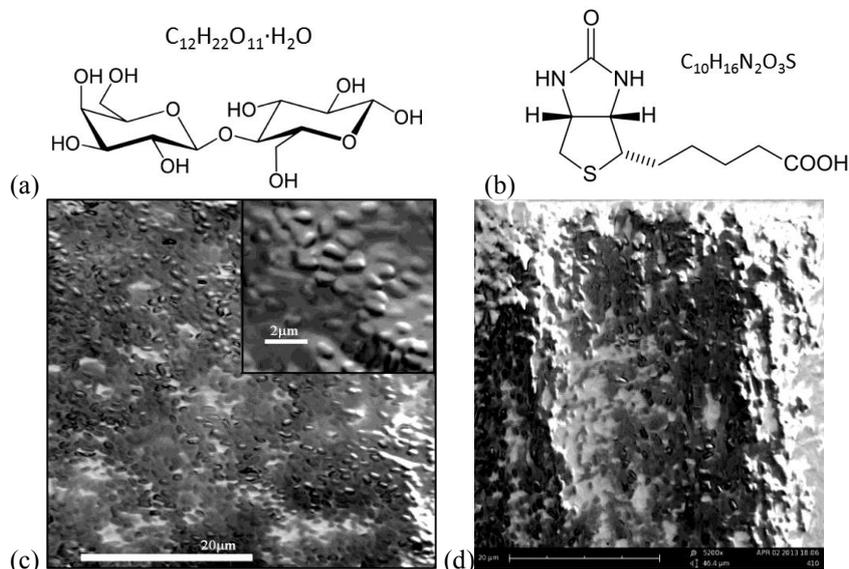

Fig. 2. Structural formula for lactose monohydrate (a) and biotin (b). (c) SEM for the B. thuringiensis (Bt) sample. (d) SEM for the B. cereus (Bc) sample. The white horizontal line at the bottom left of both photos is the 20-um scale bar. The blow-up embedded in 2(c) shows the telltale ~1.2-um-long ellipsoidal shape of the spores.

The bioparticle samples were endospores of two Bacillus species: B. thuringiensis (Bt), subspecies kurstaki, and B. cereus (Bc). Bt is a common soil bacteria and insect pathogen [10-11]. Upon sporulation, this bacteria produces bi-pyramidal protein crystals, δ-endotoxins or Cry proteins which are toxic to a variety of insects but are not harmful to humans. While Cry toxins have been widely applied as a "green" pesticide, recent use introduces Cry proteins into transgenic crops, providing a more targeted approach to insect management. The endospore and crystal are simultaneously present after an environmental stress induces sporulation in the vegetative cell, but are typically released after cell lysis. The spores can endure harsh environments such as aridity, extreme cold and heat and some radioactivity. When nutrition, temperature and other environmental conditions become favorable, spores germinate, develop into rapidly growing vegetative cells and begin another life cycle [12]. More relevant to biosensing and the CBRNE underlying theme of this NATO Workshop, Bt and Bc are in the same genus as B. anthracis [13], the notorious "Anthrax" bio-toxin. So the identification of THz signatures belonging to Bt or Bc could provide impetus for an important application of THz spectroscopic sensing, but without the handling hazards of standard biosensing techniques.

The Bacillus spores were cultured from isolated stock bacteria in 50 ml tryptic soy broth (TSB; Becton, Dickinson, Sparks, MD) at 30 C. Bacterial cultures were transferred to a test tube and centrifuged at 5000·g for 30 min. The liquid broth was removed, and then the bacteria were maintained at 4°C in the sealed test tube. Just prior to measurement, a concentrated, 0.2 mL paste-like sample of Bt or Bc bacteria (without broth) was removed from the sealed tube, and spread evenly with a spatula over ≈ 2 cm diameter on the filter paper. The thickness of the resulting samples ranged from 200 to 600 um. Scanning electron micrographs (SEMs) for the Bt and Bc pastes are shown in Figs. 2(c) and (d), respectively, each showing a large fraction of spores as proven by the zoom-in of Fig. 2(c).

Three samples of Bt were tested for the present study: sample Bt1 containing vegetative cells but negligible spores; sample Bt2 containing a mixture of vegetative cells and spores at low hydration; and sample Bt3 containing vegetative cells and spores at modest hydration. Also three samples of Bc were tested: sample Bc1 containing a mixture of vegetative spores at a hydration level comparable to Bt3; sample Bc2 the same as Bc1 but dried out thoroughly to have negligible hydration; and sample Bc3, the same as Bc2 but re-hydrated to a saturated level ("muddy" texture). Because of our lacking knowledge of the broth compositions and the degree of centrifuge concentration, the hydration levels are only qualitative and based on visual observation.

For all samples the spectrometric procedure was as follows. First, the THz beam path was blocked completely to obtain the noise floor spectrum $P_N(\nu)$. Then a blank sample holder was added to the beam path: an empty encapsulated ring for the biomolecular sample, or bare filter paper for the bioparticle sample. In both cases this yielded the background spectrum $P_B(\nu)$. Then the real samples were added to the beam path to yield the sample spectrum $P_S(\nu)$. Finally, the transmittance was calculated as $T(\nu) = [P_S(\nu)-P_N(\nu)]/[P_B(\nu)-P_N(\nu)]$. For the biomolecular samples, the time between successive scans corresponding to different hydration levels, was approximately 10 minutes.

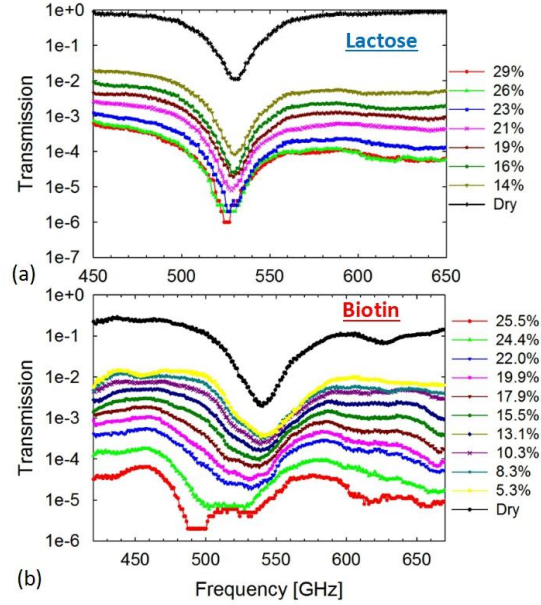

Fig. 3. Raw transmission data for two biomolecules chosen for this study: (a) lactose monohydrate, and (b) biotin.

## 1.3 Experimental Results

### 1.3.1. Biomolecules

The experimental transmission results for lactose monohydrate and biotin are displayed in Fig. 3(a) and (b), respectively. The "dry" spectrum of lactose monohydrate is in good agreement with previous, high-resolution experiments and signature analysis, showing that the signature is well fit by a Lorentzian model [14]. In the dry state of Fig. 3(a), the signature center frequency is 530 GHz and the full-width at half-maximum (FWHM) is ≈27 GHz. But when the hydration increases above ≈10%, the center frequency "red-shifts" significantly. The maximum shift is ≈4.0 GHz at 26% hydration: well-resolved by the coherent frequency-domain (photomixing) spectrometer used in this study.

For the biotin sample, the dry center frequency is ≈540 GHz with FWHM of ≈ 55 GHz - much broader than the lactose [14]. The maximum shift is ≈16 GHz at a hydration of 22%, and the shifted spectrum is significantly broadened compared to the "dry" spectrum - by approximately 50%. Upon further inspection, it appears that beyond 20% hydration in Fig. 3(b), the signature may be transforming into a "doublet", and beyond 25% may be saturating at the noise floor of the instrument.

This exemplifies a key problem with THz biospectroscopy, which is that the aqueous local environment is often critical. That is, water in moderate concentrations generally strengthens the THz signatures, particularly by inducing a polarity if one doesn't already exist or is weak. But at larger concentrations the water masks the THz signatures at least through Debye absorption. This is evident in the lactose data of Fig. 3(a) at 29.0% hydration, and in the biotin data of Fig. 3(b) at 25.5%, where the transmittance spectra saturate at the signature-center frequency because the transmitted signal drops into the instrumental noise floor.

### 1.3.2. Bioparticles

The transmission spectra of samples Bt1, Bt2, and Bt3 are plotted in Figs. 4(a), (b), and (c), respectively. Sample Bt1 having negligible sporulation and low hydration, shows a relatively high transmission but no obvious signatures except for an anomalous one near 700 GHz. Sample Bt2 having significant sporulation and also low hydration, shows four distinct signatures centered at 966, 1026, 1098, and 1164 GHz. The last two coincide with well-known and strong water-vapor absorption lines [15], but the others do not. And because Bt2 has spores but Bt1 is lacking, the 966 and 1026 signatures are attributed to the spores [16] and labelled α and β, respectively, in 4(b). There are also two broader, weaker features centered around 480 and 660 GHz that are not near any water vapor lines, and are addressed further below.

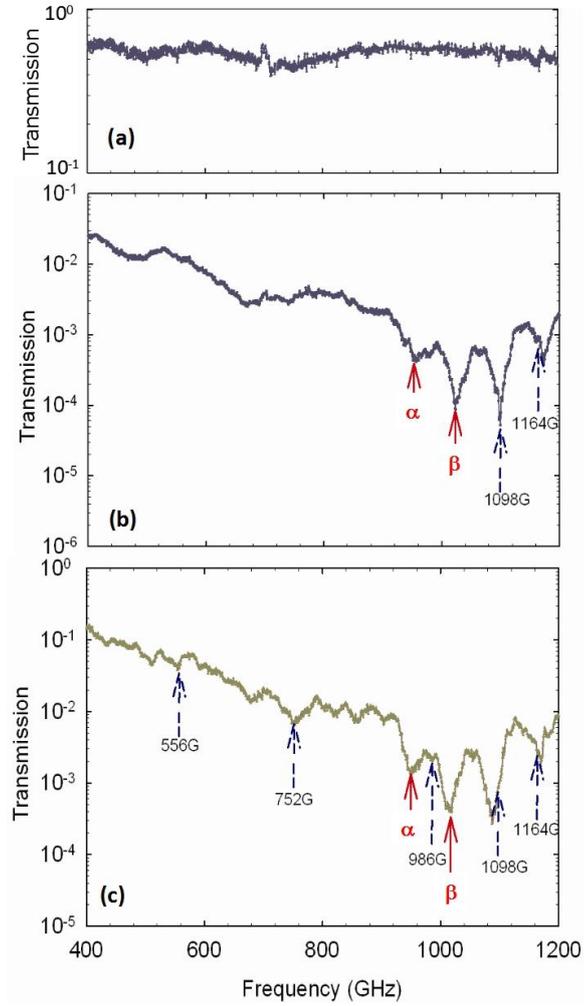

Fig. 4. (a) Transmission through Bt1 sample at low hydration and with negligible spores. (b) Transmission through Bt2 sample at low hydration with spores. (c) Transmission through Bt3 sample at moderate hydration with spores.

The transmission spectrum of samples Bt3 in Fig. 4(c) shows six distinct signatures natures centered at 556, 752, 955, 1015, 1098, and 1164.  The last two are the same water lines as above, and the first two are also well-known but weaker water lines.  The middle two are similar in shape and strength to the α and β signatures in sample Bt2, but both are red-shifted in center frequency by exactly 11.0 GHz.  Given our effective resolution of 1.0 GHz, these cannot be attributed to instrumental drift or frequency error.  So we associate them with the same spore-related signatures, α and β, and are labeled as such in Fig. 4(c)   The stationarity of the strong water-vapor lines at 1098 and 1164 GHz support this assertion, and exemplifies the utility of water vapor lines as frequency markers.

Two broader signatures are seen at 510 and 680 GHz.  It is tempting to associate these with the 480 and 660 signatures in Fig. 4(b).  However, the shift in frequency of both (30 and 20 GHz, respectively) is much greater than for the α and β signatures, and in the opposite sense (blue shift rather than red shift).  They could possibly be associated with vegetative cell material, which is supported by the presence of a broad dip around 480 GHz and the anomalous feature at 700 GHz in Fig. 4(a) (for the cell-rich, spore-sparse sample Bt1).  However, more research is necessary to strengthen this claim.

The transmission spectra of samples Bc1, Bc2, and Bc3 are plotted in Fig. 5.  Sample Bc1, known to have a significant fraction of spores, shows a predominant signature centered at 880 GHz. It is similar in shape and depth to the β signature of samples Bt2 and Bt3, but more than 100 GHz lower in frequency. The low background transmission at frequencies well below 880 GHz is consistent with the broadband (Debye) absorption expected from the hydration, and is similar to that for the hydrated Bt samples in Figs. 4(b) and (c). When Bc1 was thoroughly dried to become Bc2, the signature completely disappears and the background transmission rises dramatically, similar to sample Bt1 in Fig. 4(a). Finally, when Bc2 was re-hydrated to levels even greater than Bc1 or Bt3, the resulting Bc3 transmission dropped to the lowest levels seen with Bc2. And the signature seen in Bc1 did not reappear in any form. This suggests that Bc3 had too much hydration, likely much higher than the 25.5% where the biotin signature began to deform in Fig. 3(b). And it is strong evidence that these Bacillus spore signatures are critically dependent on hydration: with too little hydration, there is no signature like Bc2, and with too much hydration there is no signature like Bc3.

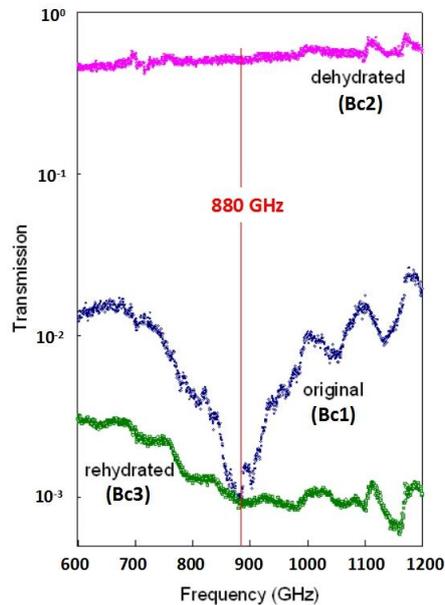

Fig. 5.  Transmission through Bc sample Bc1 (after preparation at similar hydration as Bt2), Bc2 (dehydrated form of Bc1), and Bc3 (rehydrated form of Bc2).

## 1.4 Analysis

As in most forms of spectroscopy, there is often a wealth of physical information in the lineshape of absorption signatures. Lineshape analysis has already been carried out for the biomolecules [14, 9] and bioparticles [16] studied here under dry or unknown-hydration conditions. In all cases, the line absorption coefficients α were found to be well fit by the famous Lorentzian function for dipole oscillators. The frequency dependence in the most general case is given as

$$\alpha = A \cdot \left[ \left( \omega^2 - \omega_0^2 \right)^2 + \omega^2 \gamma^2 \right]^{-1} \qquad (1)$$

where $\omega$ is the circular frequency of incident radiation, $\omega_0$ is the natural circular oscillation frequency (in the absence of damping), $\gamma$ is the relaxation frequency (i.e., damping constant), and A is a constant that depends on the density and effective charge of the dipoles, and possibly other factors depending on the type of oscillator. Analytic evaluation of the peak absorption frequency $\omega_r$ is defined implicitly through $(d\alpha/d\omega)|_{\omega_r} = 0$, which leads to

$$\omega_r = [(\omega_0)^2 - \gamma^2/2]^{1/2} \qquad (2)$$

This is monotonically decreasing with respect to $\gamma$, meaning that $\omega_r$ is always "red-shifted" with respect to the natural frequency. This is a well-known, and intuitive result from classical mechanics and electromagnetics.

Another important prediction of the Lorentzian lineshape function is the linewidth, defined generally as the "fullwidth at half-maximum" (FWHM), or $\Delta\omega$. To find this, we solve for the two values of $\omega$ in (1) at which $\alpha$ drops by -3 dB from its peak value [determined by setting $\omega = \omega_r$ in (1)], which is $\alpha_p = A[(\omega_0\gamma)^2 - \gamma^4/4]^{-1}$. For $\gamma \ll \omega_0$, this becomes $\alpha_p \approx A(\omega_0\gamma)^{-2}$, and the -3 dB frequency is found to be $\omega_{-3dB} \approx \omega_0(1 +/- \gamma/2\omega_0)$, so that $\Delta\omega \approx \gamma$.

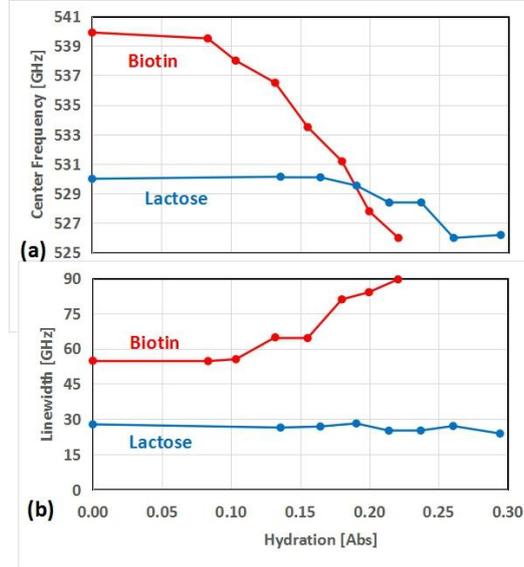

Fig. 6. (a) Experimental center frequency for the primary resonant signatures of lactose monohydrate [Fig. 3(a)] and biotin [Fig. 3(b)]. (b) Linewidth (FWHM) calculated by a Lorentzian fit of the signatures in Figs. 3(a) and (b).

Given that the Lorentzian model is known to be a good fit to the strongest biosignatures studied here, we can examine the data with a logical inquiry. The absorption peaks were found to red-shift with increasing hydration, which is also consistent with the Lorentzian model if the damping increases with hydration too. This is a reasonable physical assumption, and if true, then the linewidth should also increase with hydration as $\Delta\omega \approx \gamma$. We will investigate this next for both the biomolecular and bioparticulate signatures.

### 1.4.1. Biomolecules

Fig. 6 shows the lineshape results for the experimental lactose-monohydrate and biotin results of Figs. 3(a) and (b), respectively. The center frequency in Fig. 6(a) always red-shifts, the biotin frequency dropping rapidly above ~10% hydration, and the lactose not changing significantly until ~20%. The linewidth is obtained by a Lorentzian fitting, and the FWHM (linear frequency) is $\Delta f = \Delta\omega/2\pi = \gamma/2\pi$. The results are plotted in Fig. 6(b), showing that the biotin at zero hydration is almost twice as broad as the lactose, and then increases dramatically above ~10% consistent qualitatively with the red-shift in center frequency. In contrast, the lactose FWHM does not change significantly, even above the ~20% hydration where the center frequency drops.

To take the analysis one step further, we can use the center frequency from 6(a) to calculate the Lorentzian linewidth using a rearrangement of Eq. (2): $\Delta f = \gamma/2\pi =$ $\{2[(\omega_0)^2-(\omega_r)^2]\}^{1/2}/2\pi$. This is plotted in Fig. 7 in comparison with the experimental $\Delta f$ for lactose [7(a)] and biotin [7(b)], using $f_0 = \omega_0/2\pi$ =530.4 GHz (from lactose dry sample), and $f_0$=541.3 GHz (biotin dry sample). Clearly, the agreement for lactose is not good, the Lorentz model predicting a significant change in linewidth above ~20% hydration that is not observed experimentally. For biotin, the agreement is better qualitatively, but the Lorentz model predicts a linewidth increasing much faster with hydration than the experiment does.

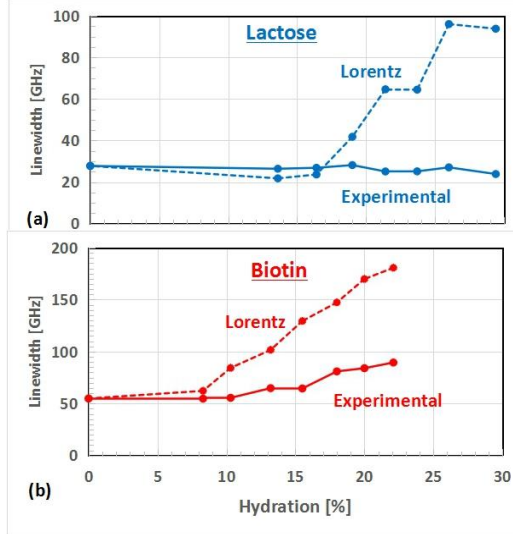

Fig. 7. Comparison of experimental linewidth (FWHM) to Lorentzian-fit linewidth for lactose (a) and biotin (b) samples vs hydration.

### 1.4.2. Bioparticles

For the bioparticles (Bacillus endospores), we cannot do such a precise analysis with the Lorentz model because we do not know $\omega_0$, the dry resonant frequency. As mentioned earlier, this is probably because hydration plays two roles with the endospores: (1) induces polarity and (2) increases damping. So no hydration means that no signatures occur in the non-polar endospores. Nevertheless, we can check for consistency with Lorentzian behavior through the experimental center frequency and linewidth for the two signature-displaying samples of Fig. 4: Bt2 and Bt3, where the hydration of Bt3 is higher than that of Bt2. Each sample displayed two distinct signatures (labeled α and β in Fig. 4) whose center frequencies and FWHM are plotted in Figs. 8(a) and (b), respectively. Both center frequencies decrease by 11 GHz with the increased hydration, consistent with Lorentzian behavior if increased hydration creates increased damping. However, the linewidth behavior in Fig. 8(b) is ambiguous. The β signature linewidth increases with hydration, consistent qualitatively with Lorentzian behavior. But the α signature linewidth decreases with hydration, similar to lactose monohydrate in Fig. 6(b), but even more opposed. This is something we are investigating further through experiment and modeling.

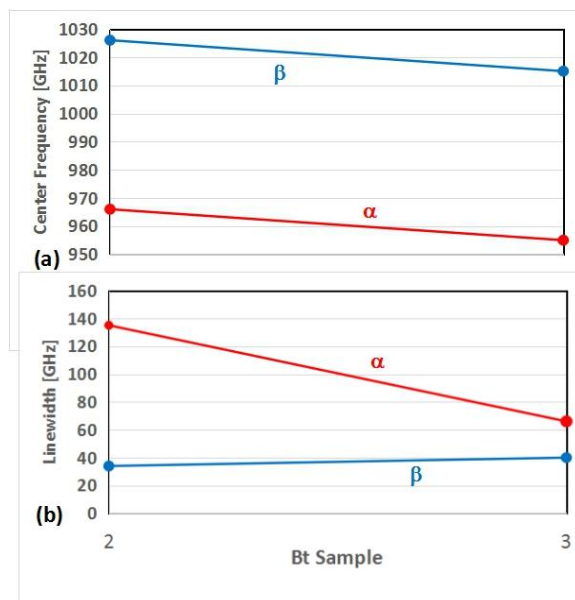

Fig. 8. (a) Center frequency of signatures α and β in samples Bt 2 and Bt3, whose hydration level is known to increase between them. (b) Linewidth (FWHM) of signatures α and β vs Bt sample hydration.

## Conclusion

We have presented the first known study of the effect of hydration on THz absorption signatures in biomolecules and bioparticles. We observe a "red-shift" in center frequency with increasing hydration in both biomolecules and bioparticles, consistent with Lorentzian-oscillator behavior. But the effect of hydration on linewidth is ambiguous, sometimes increasing the linewidth (consistent with Lorentzian behavior) and sometimes decreasing the linewidth.

This material is based upon work supported by, or in part by, the U.S. Army Research Laboratory and the U.S. Army Research Office under contract number W911NF-11-1-0024. This work was also presented in the NATO Workshop Proceedings: "THz for CBRN and Explosives Detection and Diagnosis," Pereira M., Shulika O. (eds). NATO Science for Peace and Security Series B: Physics and Biophysics. Springer, Dordrecht. https://doi.org/10.1007/978-94-024-1093-8_13